\documentclass[aoas,nameyear,seceqn,dvips]{arximspdf}
\usepackage{dcolumn}
\usepackage{graphicx}

\doi{10.1214/09-AOAS289} \volume{4} \issue{1} \pubyear{2010}
\firstpage{299} \lastpage{319}

\makeatletter
\newcolumntype{d}[1]{D{.}{.}{#1}}
\makeatother

\begin{document}
\begin{frontmatter}

\title{An MDL approach to the climate segmentation problem\protect\thanksref{T1}}
\thankstext{T1}{Supported by NSF Grants
DMS-07-07037 and DMS-09-05570 and Hong Kong Research Grant Council
Grant CERG 401507.} \runtitle{MDL climate segmentation}
\pdftitle{An MDL approach to the climate segmentation problem}
\begin{aug}
\author[a]{\fnms{QiQi} \snm{Lu}\ead[label=e1]{qlu@math.msstate.edu}},
\author[b]{\fnms{Robert} \snm{Lund}\corref{}\ead[label=e2]{lund@clemson.edu}}
\and
\author[c]{\fnms{Thomas C. M.} \snm{Lee}\ead[label=e3]{tlee@sta.cuhk.edu.hk}}
\runauthor{Q. Lu, R. Lund and T. C. M. Lee} \affiliation{Mississippi
State University, Clemson University, and Colorado State University and
Chinese University of Hong Kong}
\address[a]{Q. Lu \\
Department of Mathematics and Statistics \\
Mississippi State University\\
Mississippi State, Mississippi 39762 \\
USA\\
\printead{e1}} %adresu isvedimo komanda gale!
\address[b]{R. Lund\\
Department of Mathematical Sciences\\
Clemson University\\
Clemson, South Carolina 29634-0975 \\
USA\\
\printead{e2}}
\address[c]{T. C. M. Lee\\
Department of Statistics \\
Colorado State University\\
Fort Collins, Colorado 80523 \\
USA\\
and\\
Department of Statistics \\
Chinese University of Hong Kong \\
Shatin, Hong Kong \\
\printead{e3}}
\end{aug}

% HISTORY:
\received{\smonth{12} \syear{2008}} \revised{\smonth{9} \syear{2009}}

% ABSTRACT
%
\begin{abstract}
This paper proposes an information theory approach to estimate the
number of changepoints and their locations in a climatic time series. A
model is introduced that has an unknown number of changepoints and
allows for series autocorrelations, periodic dynamics, and a mean shift
at each changepoint time. An objective function gauging the number of
changepoints and their locations, based on a minimum description length
(MDL) information criterion, is derived. A genetic algorithm is then
developed to optimize the objective function. The methods are applied
in the analysis of a century of monthly temperatures from Tuscaloosa,
Alabama.
\end{abstract}

% KEYWORDS
%

\end{frontmatter}
%

%s1 ###
\section{Introduction}\label{sec1}

Changes in station instrumentation, location, or observer can often
induce artificial discontinuities into climatic time series. For
example, United States temperature recording stations average about six
station relocation and instrumentation changes over a century of
operation [Mitchell (\citeyear{Mitchell53})]. Many of these changepoint
times are documented in station histories; however, other changepoint
times are unknown for a variety of reasons. Even when a changepoint
time is known, one may still question whether the change instills a
mean shift in series observations. This paper proposes an information
based approach to the multiple changepoint identification
(segmentation) problem.

Our methods are specifically tailored to climatic time series in that
they allow for periodicities and autocorrelations. Multiple changepoint
detection procedures have been studied under the assumption that the
series is driven by independent and identically distributed errors
[Braun and M\"{u}ller (\citeyear{Braun98}), Caussinus and
Mestre~(\citeyear{Caussinus04}), Menne and Williams
(\citeyear{Menne05})]. This is unrealistic in climate settings where
observations display moderate to strong serial autocorrelation.
Ignoring autocorrelations can drastically alter changepoint inferences,
as positive autocorrelation can be easily mistaken for mean shifts [see
Berkes et al. (\citeyear{Berkes06}) and Lund et al.
(\citeyear{Lund07})]. Multiple changepoint methods for time series data
represent a very active area of current research [Davis, Lee, and
Rodriguez-Yam (\citeyear{Davis06}),
Fearnhead~(\citeyear{Fearnhead06})]. Series recorded daily or monthly
also display periodic dynamics. Our methods allow for seasonality by
employing a time series regression model with periodic features. In
short, this paper develops a multiple changepoint segmenter that
applies to a variety of realistic climate series.

The rest of this paper is organized as follows. Section \ref{sec2}
introduces the time series regression model that underlies our work.
Section \ref{sec3} develops an objective function for the model. The
objective function is a penalized likelihood whose penalty is based on
the minimum description length (MDL) principle. This modifies
Caussinus' and Mestre's (\citeyear{Caussinus04}) model to allow for
autocorrelation, seasonal effects, and also changes their likelihood
penalty to an MDL-based penalty. Each segment of our model is allowed
to have a distinct mean, but the autocovariance structure of each
segment is constrained to be the same. Section \ref{sec4} presents a
genetic-type algorithm capable of optimizing the objective function to
obtain estimates of the changepoint numbers, locations, and the time
series regression parameters. Section \ref{sec5} presents a short
simulation study for feel. Section \ref{sec6} applies the methods to a
century of monthly temperatures from Tuscaloosa, Alabama and Section
\ref{sec7} concludes with comments.

%s2 ###
\section{Model description}\label{sec2}

The object under study is a time series $\{ X_t \}$ governed by
periodic errors and multiple level shifts. The period of the series is
$T$ and is assumed known. The series observation during season $\nu$,
$1 \leq\nu\leq T$, of the $(n+1)$st cycle is denoted by $X_{nT+\nu}$.
The time-homogeneous and periodic notation $\{ X_t \}$ and $\{
X_{nT+\nu} \}$ are used interchangeably, the latter to emphasize
seasonality. We index the first data cycle with $n=0$ so that the first
observation is indexed by unity. For simplicity, we take $d$ complete
cycles of observations; specifically, the observed data are ordered as
$X_1, \ldots, X_N$ and $d=N/T$ is assumed a natural number.

The model driving our work is a simple linear regression in a periodic
environment:
%
%e1 ###
\begin{equation}
X_{nT+\nu} = \mu_\nu+ \alpha(nT+\nu)+ \delta_{nT+\nu}+
\varepsilon_{nT+\nu}.\label{regressioneq}
\end{equation}
In (\ref{regressioneq}) $\alpha$ is a linear trend parameter that is
assumed time homogeneous for simplicity; $\mu_\nu$ is the season $\nu$
location parameter (a detrended mean in the absence of changepoints).
The errors $\{ \varepsilon_t \}$ have zero mean and are a periodically
stationary series with period $T$ in that
%
%e2 ###
\begin{equation}
\operatorname{Cov} (\varepsilon_t, \varepsilon_s) = \operatorname{Cov}
(\varepsilon_{t+T}, \varepsilon_{s+T}) \label{defperiodic}
\end{equation}
for all integers $t$ and $s$. Many climatic series have periodic second
moments in the sense of (\ref{defperiodic}). For a sample of size $N$
with $m<N$ changepoints, the ordered times of the changepoints are
denoted by $1 < \tau_1 < \tau_2 < \cdots< \tau_m \leq N$. The number of
changepoints and the changepoint times are considered unknown. There
are $m+1$ different segments (regimes) during the observation record.
At each changepoint time, our model allows for a mean shift in the
observations. Such a structure is described by
\[
\delta_t=\cases{ \Delta_1,  &\quad $ 1 \leq t < \tau_1,$ \cr \Delta_2,
&\quad  $\tau_1 \leq t < \tau_2,$\cr \hspace*{3pt}\vdots& \quad
\hspace*{23pt}$\vdots$\cr \Delta_{m+1}, & \quad $\tau_m \leq t < N+1.$}
\label{chpt}
\]
For parameter identifiability, we take $\Delta_1 = 0$; otherwise, the
$\Delta_i$'s and $\mu_\nu$'s would become confounded. For a fixed $N$,
the mean component $E[ X_{nT+\nu}]$ in (\ref{regressioneq}) depends on
the $T+1+m$ parameters $\mu_1, \ldots, \mu_T$, $\alpha$, and $\Delta_2,
\ldots, \Delta_{m+1}$. Generalizations of (\ref{regressioneq}) are
mentioned in Section \ref{sec3} when we derive MDL codelengths.

To describe the time series component $\{ \varepsilon_{nT+\nu} \}$, we
use a causal periodic autoregression of order $p$
$[\operatorname{PAR}(p)]$. Such errors are the unique (in mean square)
solution to the periodic linear difference equation
%
%e3 ###
\begin{equation}
\varepsilon_{nT+\nu}=\sum_{k=1}^p \phi_k(\nu) \varepsilon_{nT+\nu
-k}+Z_{nT+\nu}. \label{PAR}
\end{equation}
Here, $\{ Z_t \}$ is zero mean periodic white noise with variance
$\sigma^2(\nu)$ during season $\nu$. Solutions to (\ref{PAR}) are
indeed periodic with period $T$ in the sense of (\ref{defperiodic}).
PAR models are dense in the set of short memory periodic time series
and parsimoniously describe many such series; explicit expressions for
many time series quantities are available for PARs.

In many applications, reference series are available. A reference
series is a series of the same genre as the series to be studied (the
target series) that serves to aid changepoint identification. For
example, with the Tuscaloosa temperatures examined later, series from
nearby Greensboro AL, Selma AL, and Aberdeen MS are available over the
same period of record. By constructing a target minus reference
difference series, mean shifts induced by changepoints are sometimes
illuminated. When the reference series is highly positively correlated
with the target series, the target minus reference series will have
smaller autocorrelations than the target series at all lags (this
happens when the target and reference series have the same periodic
autocovariance structure and the correlation between these two series
exceeds $1/2$ at all times). Also, the linear trend assumption is
typically more plausible for target minus reference differences than
the target series, as long-memory and other nonlinear features can be
eliminated in the subtraction. Moreover, the seasonal mean cycle is
frequently reduced or altogether eliminated in target minus reference
series. Drawbacks with reference series lie with additional
undocumented changepoints that the reference series may introduce.
Algorithms aimed at resolving which series among target and multiple
references is responsible for any found changepoints are now available
[see Menne and Williams (\citeyear{Menne05,Menne09})], but these works
do not consider seasonal features or autocorrelated errors.

Note that the difference of two series governed by (\ref{regressioneq})
again lies in (\ref{regressioneq}). Hence, in the next three sections
we simply consider a single series satisfying~(\ref{regressioneq}).
Reference series will return in Section \ref{sec6}.

The parameters in the model will become important later. The
$\operatorname{PAR}(p)$ model, including the $T$ white noise variance
parameters, has $(p+1)T$ autocovariance parameters. For a fixed $m$,
there are also the changepoint times $\tau_1, \ldots, \tau_m$ and the
mean shifts $\Delta_2, \ldots, \Delta_{m+1}$. Finally, a trend
component $\alpha$ and the seasonal means $\mu_1, \ldots, \mu_T$ are
present. Hence, given $p$ and $m$, there are $2m+1+(p+2)T$ model
parameters. Given $p$ and $m$, we will need to estimate $\tau_1,
\ldots, \tau_m$, $\Delta_2, \ldots, \Delta_{m+1}$, $\mu_1,
\ldots,\mu_T$, $\alpha$, and all $\operatorname{PAR}(p$) parameters.
Developing and optimizing an objective function for this purpose will
be the subject of our next two sections.

Before leaving the model description, we make a comment. The model
studied here allows for process changes at the changepoint times in the
form of level shifts. This is reasonable in climate cases [Vincent
(\citeyear{Vincent98}), Menne and Williams (\citeyear{Menne05}), Lund
et al. (\citeyear{Lund07})]. In other applications such as speech
recognition and finance, it may be more realistic to keep mean process
levels fixed and allow the time series parameters to change at each
changepoint time [see Inclan and Tiao (\citeyear{Inclan94}), Chen and
Gupta (\citeyear{Chen97}), Davis, Lee, and Rodriguez-Yam
(\citeyear{Davis06})].
%s3 ###
\section{An MDL objective function}\label{sec3}

To fit the above model, estimates of the changepoint numbers and
locations, as well as the model parameters, are needed. Since different
changepoint numbers refer to models with a different number of
parameters, the model dimension will also need to be estimated. This is
a~model selection problem. Popular approaches to model selection
problems include AIC (Akaike Information Criterion), BIC (Bayesian
Information Criterion), cross-validation type methods, and MDL methods.
For problems that involve the detection of regime changes, MDL methods
often provide superior empirical results [e.g., Lee
(\citeyear{Lee00,Lee02}), Davis, Lee, and Rodriguez-Yam
(\citeyear{Davis06})]. This superiority is likely due to the fact that
both AIC and BIC place the same penalty on all parameters, regardless
of the nature of the parameter (e.g., mean shift magnitudes and
changepoint times receive the same penalty). On the other hand, MDL
methods can situationally tailor penalties for parameters of different
natures, thereby accounting for whether the parameter is real or
integer-valued.

The MDL principle was developed by Rissanen
(\citeyear{Rissanen89,Rissanen07}) as a general method for solving
model selection problems. It has roots in coding and information
theories. In brief, MDL defines the best fitting model as the one that
enables the best compression of the data; for the current problem, the
data are the observed~$\{X_t\}$. There exist several versions of MDL;
the so-called two-part MDL is used here. For introductory MDL material,
see Hansen and Yu (\citeyear{Hansen01}) and Lee~(\citeyear{Lee01}).

The rest of this section develops a two-part MDL objective function for
fitting a good model. The main idea behind the two-part MDL is
described as follows. First, the data $\{X_t\}$ is decomposed into two
parts, the fitted candidate model and its corresponding residuals. MDL
methods then calculate the total codelength (i.e., the amount of
computer memory) required for storing both parts as a sum of the
codelength of the two parts. Finally, MDL methods define the best
fitting model as one that produces a minimal codelength. Intuition
behind MDL methods lies with why minimum codelength models are also
good statistical models. Essentially, it is that both good compression
and good statistical models are capable of capturing regularities in
the data.

To proceed, let $\operatorname{CL}(z)$ denote the codelength of the
object $z$. Also write a candidate fitted model as $\hat{\mathcal{M}}$
and its residuals as $\{\hat{\varepsilon}_t\}$. The codelength is
additive in that
%
%e4 ###
\begin{equation}
\operatorname{CL}( \{ X_t \})= \operatorname{CL}(\hat{\mathcal{M}}) +
\operatorname{CL}( \{ \hat{\varepsilon}_t \}). \label{decomp1}
\end{equation}
The term $\operatorname{CL}(\hat{\mathcal{M}})$ in (\ref{decomp1}) can
be viewed as a model complexity term, while $\operatorname{CL}(\{
\hat{\varepsilon}_t \} )$ can be viewed as a data fidelity term. Our
next task is to obtain a computable expression for
$\operatorname{CL}(\{X_t\})$ that can be minimized. We begin with the
calculation of $\operatorname{CL}(\hat{\mathcal{M}})$.

An important result of Rissanen (\citeyear{Rissanen89}) is that the
maximum likelihood estimate of a real-valued parameter computed from a
series of $N$ observations ($N$ is large) can be effectively encoded
with $\log_2(N)/2$ bits. The trend parameter $\alpha$ hence requires
$\log_2(N)/2$ bits to encode. The seasonal mean parameters $\mu_\nu$
are effectively estimated via seasonal sample means, each of which
contributes $\log_2(d)/2$ bits to the codelength. Given values of the
changepoint times $\tau_1, \ldots, \tau_m$, the mean shift parameter
$\Delta_j$ can be estimated with data from the $j$th segment only.
Hence, $\Delta_j$ requires $\log_2(\tau_j-\tau_{j-1})/2$ bits to encode
for $2 \leq j \leq m+1$ ($\tau_{m+1}=N+1$ is taken as a convention).
Hence, the portion of the codelength from mean parameters in the time
series regression (i.e., $\alpha, \{ \mu_\nu\}_{\nu=1}^{T}$, and $\{
\Delta_j\} _{j=2}^{m+1}$) is
%
%e5 ###
\begin{equation}
\frac{\log_2(N)}{2}+ \frac{T\log_2(d)}{2}+
\frac{1}{2}\sum_{j=2}^{m+1}\log_2(\tau_j-\tau_{j-1}). \label{piece1}
\end{equation}

The $\operatorname{PAR}(p)$ time series parameters [$\phi_k(\nu)$ for
$1 \leq k \leq p; 1 \leq\nu\leq T$ and $\sigma^2(\nu)$ for $1
\leq\nu\leq T$] are also real valued. Because $\{ \varepsilon_t \}$ is
a zero mean process, we need only consider the zero mean version of
this model. In this case, the PAR parameters can be estimated in an
efficient manner via seasonal versions of the Yule--Walker equations
[see Pagano (\citeyear{Pagano78})]. The necessary equations for this
task are presented in Shao and Lund (\citeyear{Shao04}). Yule--Walker
PAR parameter estimators are asymptotically most efficient [Pagano
(\citeyear{Pagano78})]; in fact, these estimators are the likelihood
estimators except for the edge-effects (i.e., the likelihood is
conditional on the first $p$ observations). The Yule--Walker estimators
can be computed from the sample autocovariances $\gamma_\nu(h)$ over
the lags $h =0, \ldots, p$. The lag $h$ sample autocovariance at season
$\nu$ is defined as $\hat{\gamma}_\nu(h)=d^{-1}\sum_{n=0}^{d-1}
\varepsilon_{nT+\nu} \varepsilon_{nT+\nu-h}$, where $\varepsilon_t$ is
taken as zero should a $t \leq 0$ be encountered in the summation.
Observe that $\hat{\gamma}_\nu (0)$ is a~function of $d$ series
observations for each fixed $\nu$. Moreover, $\hat{\gamma}_\nu(h)$ is
essentially computed from $2d$ observations. Hence, the total
codelength from $\operatorname{PAR}(p)$ parameters is
%
%e6 ###
\begin{equation}
\frac{T\log_2(d)}{2} + \frac{pT\log_2(2d)}{2}. \label{piece2}
\end{equation}

The parameters $\tau_1, \ldots, \tau_m$ are integers and must be
treated as such. Arguing as in Davis, Lee, and Rodriguez-Yam
(\citeyear{Davis06}), an integer parameter bounded by~$Q$ takes
$\log_2(Q)$ bits to encode. Since the $\tau_j$'s are ordered, we have
$\tau_j < \tau_{j+1}$. This differs from Davis, Lee, and Rodriguez-Yam
(\citeyear{Davis06}) in that we do not loosely bound
$\tau_j-\tau_{j-1}$ by $N$ for each $j$. In short, the codelength
induced by the changepoint times that we use is
%
%e7 ###
\begin{equation}
\sum_{j=2}^{m} \log_2(\tau_j)+ \log_2(N). \label{piece3}
\end{equation}

Finally, the model orders $p$ and $m$ contribute
%
%e8 ###
\begin{equation}
\log_2(p)+\log_2(m) \label{piece4}
\end{equation}
bits to the codelength. While $m$ is bounded by $N$, typical values of
$m$ are significantly smaller than $N$ and a penalty of $\log_2(N)$
would be too much for $m$ changepoints.

Adding (\ref{piece1})--(\ref{piece4}) gives
%
%e9 ###
\begin{eqnarray}\label{comp1}
\operatorname{CL}(\hat{\mathcal{M}})&=& \frac{3}{2} \log_2(N)+ T
\log_2(d) + \frac{1}{2}\sum_{j=2}^{m+1} \log_2(\tau_j - \tau_{j-1})
\nonumber \\[-8pt]
\\[-8pt]
\nonumber
 &&{} + \frac{pT \log_2(2d)}{2}+ \sum_{j=2}^m \log_2(\tau_j) +
\log_2(m) +\log_2(p).
\end{eqnarray}

Moving to $\operatorname{CL}(\{ \hat{\varepsilon_t} \})$, a fundamental
result of Rissanen (\citeyear{Rissanen89}) is that this quantity equals
the negative logarithm (base 2) of the likelihood of the fitted model
$\hat{\mathcal{M}}$. For the present problem, this conditional
likelihood can be calculated as follows. A Gaussian joint density of
observations from the model, denoted by $L$, takes the classical
innovations form modified to allow for series periodicities and level
shifts at the changepoint times:
%
%e10 ###
\begin{equation}
L= (2 \pi)^{-N/2} \Biggl( \prod_{t=1}^N v_t \Biggr)^{-1/2} \exp\Biggl[
-\frac{1}{2} \sum_{t=1}^N \frac{(X_t -\hat{X}_t)^2}{v_t} \Biggr].
\label{rawlkhd}
\end{equation}
Here, $\hat{X}_t=P(X_t|X_1, \ldots, X_{t-1}, 1)$ is the best
one-step-ahead predictor of $X_t$ from linear combinations of a
constant and $X_1, \ldots, X_{t-1}$. Also, $v_t =E[ (X_t
-\hat{X}_t)^2]$ is the mean squared error (unconditional) of the
one-step-ahead predictor.

The one-step-ahead prediction equations and mean squared errors for the
$\operatorname{PAR}(p)$ setup are easily expressed:
\[
\hat{X}_{nT+\nu}= E[ X_{nT+\nu} ] +\sum_{k=1}^p
\phi_k(\nu)(X_{nT+\nu-k}- E[ X_{nT+\nu-k} ]),\qquad      nT+\nu> p,
\]
where $E[ X_{nT+\nu} ]=\mu_\nu+ \alpha(nT+\nu) + \delta_{nT+\nu}$ is
the mean function. Computing $\hat{X}_t$ and $v_t$ for $t \leq p$ is
done as in Shao and Lund (\citeyear{Shao04}). Taking a negative
logarithm in (\ref{rawlkhd}) gives
%
%e11 ###
\begin{equation}
\qquad \operatorname{CL}( \{ \hat{\varepsilon}_t \})= \frac{N}{2} \log_2( 2
\pi) + \frac{1}{2} \sum_{t=1}^N \log_2(v_t) +\frac{1}{2} \log_2(e)
\sum_{t=1}^N \frac{ (X_t -\hat{X}_t)^2}{v_t}. \label{comp2}
\end{equation}

Substituting (\ref{comp1}) and (\ref{comp2}) into (\ref{decomp1}), we
arrive at the following approximation:
\begin{eqnarray*}
\operatorname{CL}(\{X_t\})&=& \log_2(e) \Biggl[ \frac{3}{2} \ln(N)+ T
\ln (d) + \frac{1}{2} \sum_{j=2}^{m+1} \ln(\tau_j - \tau_{j-1})+
\frac{pT\ln(2d)}{2}\\
&&{}\hspace*{35pt}+\sum_{j=2}^m \ln(\tau_j)+ \ln(m)+\ln(p)+ \frac{N}{2} \ln(2\pi)\\
&&{}\hspace*{128pt} + \frac{1}{2} \sum_{t=1}^N \ln(v_t) +\frac{1}{2}
\sum_{t=1}^N \frac{ (X_t -\hat{X}_t)^2}{v_t} \Biggr].
\end{eqnarray*}
Because $N$, $d$, and $T$ are constant, our objective function for the
model ${\mathcal{M}}$, denoted by $\operatorname{MDL}({\mathcal{M}})$,
can be taken as
%
%e12 ###
\begin{eqnarray}\label{MDL1}
\operatorname{MDL}({\mathcal{M}})&=& \frac{1}{2} \sum_{j=2}^{m+1}
\ln(\tau_j - \tau_{j-1})+ \frac{pT\ln(2d)}{2}+\sum_{j=2}^m \ln(\tau_j)
\nonumber \\[-8pt]
\\[-8pt]
\nonumber
 &&{} + \ln(m) + \ln(p)+ \frac{1}{2} \sum_{t=1}^N \ln(v_t) +
\frac{1}{2} \sum_{t=1}^N \frac{ (X_t -\hat{X}_t)^2}{v_t}.\nonumber
\end{eqnarray}

MDLs for variants of the model in (\ref{regressioneq}) are worth
mentioning. Should one also allow the trend to change with each regime,
the codelength becomes, after appropriate
modification of (\ref{piece1}),
\begin{eqnarray*}
\operatorname{MDL}({\mathcal{M}})&=& \sum_{j=1}^{m+1} \ln(\tau_j -
\tau_{j-1}) - \ln(\tau_1-1)/2+ \frac{pT\ln(2d)}{2}+ \sum_{j=2}^m
\ln(\tau_j) \\
&&{}+\ln(m) + \ln(p)+ \frac{1}{2} \sum_{t=1}^N\ln(v_t) + \frac{1}{2}
\sum_{t=1}^N \frac{ (X_t -\hat{X}_t)^2}{v_t},
\end{eqnarray*}
where $\tau_0=1$ is taken as a convention. If the seasonal
location parameters $\mu_\nu$ are consolidated to a single $\mu$, then
an appropriate MDL (this assumes a single trend parameter) is
%
%e13 ###
\begin{eqnarray}
\operatorname{MDL}({\mathcal{M}})&=& \frac{1}{2} \sum_{j=1}^{m+1}
\ln(\tau_j - \tau_{j-1})+ \frac{pT\ln(2d)}{2}+\sum_{j=2}^m \ln(\tau_j)\nonumber \\[-8pt]
\\[-8pt]
\nonumber
 &&{}+
\ln(m) + \ln(p) + \frac{1}{2} \sum_{t=1}^N \ln(v_t) + \frac{1}{2}
\sum_{t=1}^N \frac{ (X_t-\hat{X}_t)^2}{v_t},\nonumber
\end{eqnarray}
which is (\ref{MDL1}) expect for the
$2^{-1}\ln(\tau_1-\tau_0)$ term added in the first summation. MDLs for
models where the structural form of the regression changes segment by
segment are harder to quantify, but also have climate ramifications and
are currently being investigated.

By an MDL model, we refer to a model $\hat{{\mathcal{M}}}$ that
minimizes a MDL score over the class of models being considered.
Practical minimization of $\operatorname{MDL}({\mathcal{M}})$ over all
admissible models is not a trivial task, which brings us to our next
section.
%s4 ###
\section{Optimizing the objective function}\label{sec4}

First, suppose that we know $p$ and $m$ and the changepoint times
$\tau_1, \ldots, \tau_m$. Then computation of
$\operatorname{MDL}(\hat{\mathcal{M}})$ proceeds as follows.
Computation of the model codelength given the parameters is
straightforward. For computation of the likelihood contribution to the
codelength, write (\ref{regressioneq}) in the general linear models
form
%
%e14 ###
\begin{equation}
\vec{X}= D \vec{\beta} + \vec{\varepsilon}. \label{GLM}
\end{equation}
In (\ref{GLM}), $\vec{\beta}=(\mu_1, \ldots, \mu_T, \alpha, \Delta_2,
\ldots, \Delta_{m+1})^\prime$, $\vec{X}=(X_1, \ldots, X_N)^\prime$,
$\vec{\varepsilon}=(\varepsilon_1, \ldots,\break \varepsilon
_N)^\prime$, and $D$ is the $N \times(T+1+m)$ design matrix
\[
D=[ S | C | R],
\]
where $S$ is an $N \times T$ dimensional seasonal indicator
matrix (all entries are zero except $S_{t,\nu}=1$ if $t=\ell T+ \nu$
for some $\ell\in\{ 0, \ldots, d-1 \}$), $C$ is an $N \times1$ vector
with $C_t=t$, and $R$ is an $N \times m$ dimensional matrix with all
zero entries except $R_{t,j}=1$ when time $t$, $1 \leq t \leq N$, is
observed during regime $j$ for $2 \leq j \leq m+1$.

We first estimate $\vec{\beta}$ with ordinary least squares methods.
From the estimated~$\vec{\beta}$, residuals of this model fit are next
computed. From these residuals and a PAR order parameter $p$, estimates
of $\phi_k(\nu)$ for $1 \leq\nu\leq T$ and $1 \leq k \leq p$ and
$\sigma^2(\nu)$ for $1 \leq\nu\leq T$ are constructed via seasonal
Yule--Walker moment estimation methods. With estimates of the
$\phi_k(\nu)$'s and $\sigma^2(\nu)$'s, one can return to (\ref{GLM})
and compute generalized weighted least squares estimators of
$\vec{\beta}$. New residuals are computed and the process is iterated
in a Cochrane--Orcutt fashion [see Cochrane and Orcutt
(\citeyear{Cochrane49})] until convergence is achieved. The process
gives jointly optimal estimators of $\vec{\beta}$ and the
$\phi_k(\nu)$'s and $\sigma^2(\nu)$'s. Typically, only several
iterations are needed.

The above enables us to quickly compute a codelength for fixed values
of $p$,~$m$, and $\tau_1, \ldots, \tau_m$. However, not counting
different values of $p$, there are $2^N$ different configurations of
$m$ and $\tau_1, \ldots, \tau_m$ that must be considered. In other
words, the parameter space has a huge cardinality. To optimize the
codelength over this parameter space, we now introduce a genetic
algorithm.

A genetic algorithm (GA) is a stochastic search that can be applied to
a variety of combinatorial optimization problems [Goldberg
(\citeyear{Goldberg89}), Davis (\citeyear{Davis91}),
Reeves~(\citeyear{Reeves93})]. The basic principles of GAs were first
developed by Holland~(\citeyear{Holland75}) and are designed to mimic
the genetic process of natural selection and evolution. GAs start with
an initial population of individuals, each representing a possible
solution to the given problem. Each individual or \textit{chromosome}
in the population is evaluated to determine how well it scores with
respect to the objective function. Highly fit individuals are more
likely to be selected as parents for reproduction. In a
\textit{crossover} procedure, the \textit{offspring}
(\textit{children}) share some characteristics of the parents.
\textit{Mutation} is often applied after crossover to introduce random
changes to the current population with a small probability. Mutation
increases population diversity. The offspring are used to construct a
new generation by either a generational approach (replacing the whole
population) or a steady-state approach (replacing a few of the less fit
individuals). This process is repeated until an individual is found
that roughly optimizes the objective function.

The GA used in this study is described as follows.

\textit{Chromosome representation}: The first step in designing a GA is
to create a suitable chromosome representation for the problem. Here,
any individual (model) can be described as a set of parameters: the
number of changepoints $m$, the order of the PAR model $p$, and the
changepoint locations $\tau_1, \ldots, \tau_m$. Once these parameters
are fixed, the regression parameters in the model (\ref{regressioneq})
can be estimated using the methods described above. Hence, the
chromosome, denoted by $u=(m, p, \tau_1, \ldots, \tau_m)$, is an
integer vector of length $m+2$. The lengths of the chromosomes in the
population depend on the number of changepoints. A minimum number of
observations in each regime is set to $m_\ell T$ to ensure that
reasonable mean shift estimates are obtained in all segments. Here,
$m_\ell$ is the minimum number of cycles between adjacent changepoints.
Our work will take $m_\ell=1$ (no changepoints within a year for
monthly data). Also, we impose the upper bound $p_{\rm max}$ for the
order of the $\operatorname{PAR}(p)$ model; $p_{\rm max}=3$ is used in
the forthcoming simulation study and examples.

\textit{Initial population generation}: For each individual, the PAR
order $p$ is first randomly selected with equal probabilities from the
set $\{0,1,2,3\}$. The changepoint numbers $m$ and locations are then
independently simulated as follows. There is a probability $p_b$,
essentially representing the probability that any admissible time is
selected as a changepoint. Since there can be no changepoint before
time $t=1+m_\ell T$, we first examine time $t=1+m_\ell T$, flipping a
coin with heads probability $p_b$. If the flip is heads, $t=1+m_\ell T$
is declared to be the first changepoint ($\tau_1=1+m_\ell T$) and
attention shifts to the next possible changepoint time, which is time
$1+2m_\ell T$. But if the flip is tails, $t=1+m_\ell T$ is not chosen
as a changepoint and we move to the next location at $t=1+m_\ell T+1$,
independently flipping the coin again. The process is continued in a
similar manner until the last admissible changepoint time at
$t=N-m_\ell T$ is exceeded. The population size $n_p= 30$ is used in
this study and $p_b$ is set to be $0.06d$ (six changepoints over a
century).

\textit{Crossover}: Pairs of parent chromosomes, representing mother
and father, are randomly selected from the initial population or
current population by a linear ranking/selection method. That is, a
selection probability is assigned to an individual that is proportional
to the individual's rank in optimizing the objective function. The
least fit individual is assigned the rank $0$ and the most fit
individual is assigned the rank $n_p-1$. A crossover procedure, as
explained in the paragraph below, is then applied to the parents to
produce offspring for the next generation. The probability that any two
parents have children, denoted by $p_c$, is set to $p_c=1-m_\ell/d$.

In our GA implementation, only one pair of parent chromosomes is chosen
from the current generation and one child is produced by ``mixing'' two
parent chromosomes with a uniform crossover. This works as follows. The
child's PAR order $p$ is either the mother's or the father's PAR order,
with both being equally likely. The child's changepoint locations are
randomly selected using all admissible changepoint locations from
\textit{both} mother and father. For example, for $N=1200$, $T=12$, and
$m_\ell=1$, suppose the mother's chromosome has 3 changepoints at the
times $t=200, 320, 600$ and the father's chromosome has 4 changepoints
at the times $t=205, 300, 710, 850$. First, all changepoints from
mother and father are mixed together and sorted from smallest to
largest, yielding the string $(200, 205, 300, 320, 600, 710, 850)$. We
select the first changepoint of the child at $t=200$ with probability
$0.5$. If $t=200$ is selected as a changepoint, then we discard the
changepoint $t=205$ (it would violate segmentation spacing
requirements) and move to the next candidate changepoint at $t=300$,
again doing a fifty-fifty selection/inclusion randomization. If $t=200$
is not chosen as one of the child's changepoints, we move to the next
changepoint at $t=205$ with the same fifty-fifty selection criterion.
The child's $m$ is simply the number of retained changepoints.

\textit{Mutation}: Mutation is applied to the child after crossover
with a constant probability $p_m$. The probability $p_m$ is typically
low; we use $p_m=0.05$ in the following examples. The PAR order $p$ for
the new chromosome produced by mutation is equal to the child's $p$
with a probability of $0.5$. Then changepoint locations can either take
on the corresponding changepoints from the child's or be a new set
randomly selected from the parameter space. Mutation ensures that no
solution in the admissible parameter space has a zero probability of
being examined.

\textit{New generation}: The steady-state replacement method with a
duplication check as suggested by Davis (\citeyear{Davis91}) is applied
here to form a new generation. One advantage of the steady-state
approach over the generational approach is that it typically finds
better solutions faster. In our implementation of the steady-state
approach, only one individual is replaced in the current generation by
a child after crossover and/or mutation. This allows parents and
offsprings to live concurrently, which is true for long-lived species
[Beasley, Bull, and Martin (\citeyear{Beasley93})]. If the child is
already present in the current generation, this child will be discarded
and another child must be produced by the selection-crossover-mutation
process. The duplication check is applied to all new children until a
child is found that is not present in the current generation. In this
way, duplicate solutions and premature convergence are significantly
avoided.

\textit{Migration}: Migrations act to speed up convergence of the GA
and can be implemented via a parallel scheme [Davis
(\citeyear{Davis91}), Alba and Troya (\citeyear{Alba99})]. Migration
also reduces the probability of premature convergence. The population
is divided into several different sub-populations (islands). Highly fit
individuals periodically migrate between the islands. The island model
GA is controlled by several parameters, such as the number of islands
$N_I$, the frequency of migration $M_i$, the number of migrants $M_n$,
and the method used to select which individuals migrate. The migration
policy used here is as follow. After every $M_i$ generations, the least
fit individual on island $j$, $j=1, \ldots, N_I$, is replaced by the
best individual on island~$i$, which is randomly selected among all
other islands ($j \ne i$). Therefore, each island sends and receives
individuals from different islands throughout the duration of the
search process. Here, we set $N_I=40$, $M_i=5$, and $M_n=1$.

\textit{Convergence and stopping criteria}: We follow the criterion of
Davis, Lee, and Rodriguez-Yam (\citeyear{Davis06}) to declare
convergence and terminate the GA. If the overall best individual at the
end of each migration does not change for $M_c$ consecutive migrations,
then the GA is deemed to have converged to this best individual.
Additionally, if the total number of migrations exceeds a predetermined
maximum number $M^*$, then the search process is terminated and the
best individual in the~$M^*$th migration is taken as the optimal
solution to the given problem. The parameters~$M_c$ and $M^*$ are taken
as 10 and 25 in the study, respectively.
%s5 ###
\section{A simulation study}\label{sec5}

This section investigates the accuracy of the above methods via
simulation. This study is designed to correspond to the simulation
study in Caussinus and Mestre (\citeyear{Caussinus04}). Elaborating, we
will simulate a thousand series and apply our methods to each series.
Each series contains a century ($d=100$) of monthly data ($T=12$) with
six ($m=6$) changepoints. This corresponds to the average number of
changepoints over a century of operation reported in
Mitchell~(\citeyear{Mitchell53}). The changepoint mean shifts in every
series occur at the times $\tau_1=240$, $\tau_2=480$, $\tau_3=600$,
$\tau_4=840$, $\tau_5=900$, and $\tau_6=1020$. The error terms~$\{
\varepsilon_t \}$ are simulated as a Gaussian first order periodic
autoregression ($p=1$) with parameters $\phi_1(\nu)$ and
$\sigma^2(\nu)$ as specified in Table \ref{tab1}; the seasonal means
$\mu_\nu$ are also listed in Table \ref{tab1} and are in degrees
Celsius. These values are those that were estimated for 50 years of
monthly temperatures from Longmire, Washington, which was studied in
Lund et al. (\citeyear{Lund07}). The trend parameter $\alpha$ was set
to zero in all simulations.

%
%t1 ###
\begin{table}[t]
\caption{Simulation parameters}\label{tab1}
\begin{tabular*}{6cm}{@{\extracolsep{\fill}}ld{2.2}d{2.3}c@{}}
\hline
$\bolds{\nu}$ & \multicolumn{1}{c}{$\bolds{\mu_\nu}$} & \multicolumn{1}{c}{$\bolds{\phi_1(\nu)}$} & $\bolds{\sigma^2(\nu)}$ \\
\hline
\phantom{0}1 & -0.61 & 0.272 & 2.713 \\
\phantom{0}2 & 0.99 & 0.284 & 2.748 \\
\phantom{0}3 & 2.35 & 0.478 & 1.871 \\
\phantom{0}4 & 4.91 & 0.286 & 1.717 \\
\phantom{0}5 & 8.74 & 0.335 & 2.474 \\
\phantom{0}6 & 12.15 & 0.279 & 2.403 \\
\phantom{0}7 & 15.51 & 0.245 & 2.569 \\
\phantom{0}8 & 15.47 & 0.137 & 1.910 \\
\phantom{0}9 & 12.79 & -0.127 & 2.826 \\
10 & 7.82 & 0.082 & 2.488 \\
11 & 2.32 & 0.196 & 2.394 \\
12 & -0.25 & 0.214 & 2.256 \\
\hline
\end{tabular*}
\end{table}

The magnitude of the mean shifts $\Delta_2, \ldots, \Delta_7$ are
critical. Big mean shifts make changepoints easier to detect. To
facilitate interpretability, we use a common mean shift magnitude
$\Delta>0$ at all changepoint times. For instance, if the current
regime has mean level $c$ (trend and seasonal effects are assumed zero
here), the next regime will have mean $c+\Delta$ or $c-\Delta$, with a
fifty-fifty chance of shifting up or down at each changepoint time. It
follows that $\Delta=|\Delta_j-\Delta_{j-1}|$ for $j=2, \ldots, 7$.

\begin{figure}

\includegraphics{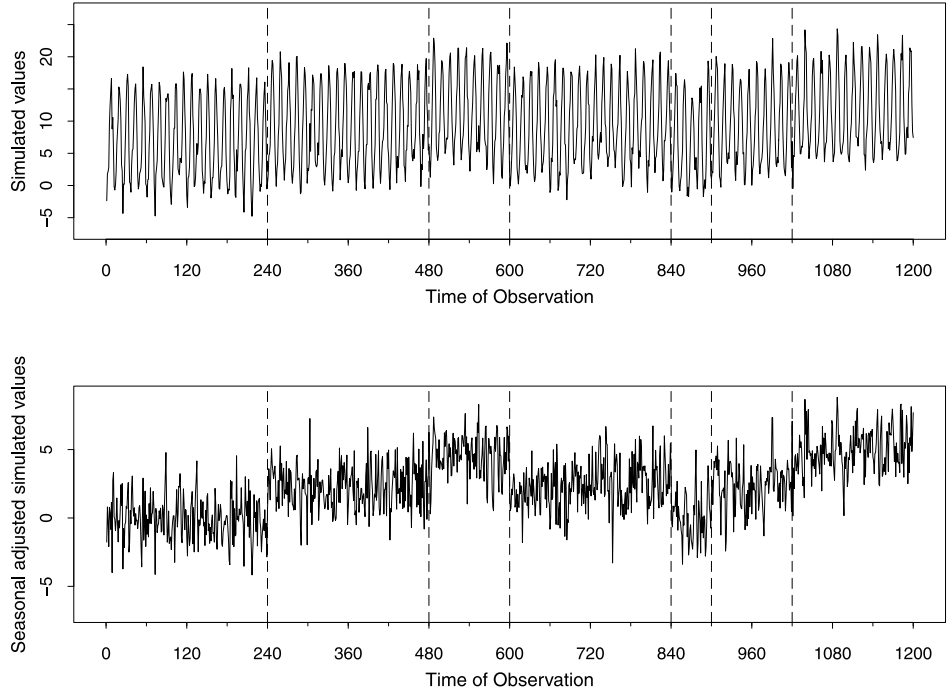}

\caption{A simulated series with six
changepoints.}\label{fig1}\vspace*{-3pt}
\end{figure}

The ability of our model to detect mean shifts can be roughly
quantified by the mean shift magnitude relative to the process standard
deviation (the latter averaged over a complete seasonal cycle). A
parameter quantifying such aspects, denoted by~$\kappa$, is
\[
\kappa= \frac{\Delta}{\sqrt{T^{-1}\sum_{\nu=1}^T
\operatorname{Var}(\varepsilon_{nT+\nu})}}.
\]
Better quantifiers of changepoint detection power may well exist, but
derivation of such quantities would be difficult and is tangential to
our points. Below, we consider three different $\kappa$ values: 1.0,
1.5, and 2.0. The larger $\kappa$ is, the easier it is to detect
changepoints. A realization of a temperature series with $\kappa=1.5$
is plotted in Figure \ref{fig1} for feel.

%
%t2 ###
\begin{table}[b]\vspace*{-2pt}
\tablewidth=6cm \caption{Estimated changepoint numbers and PAR(1) order
when $m=6$}\label{tab2}
\begin{tabular*}{6cm}{@{\extracolsep{\fill}}ld{3.3}d{2.3}d{2.3}@{}}
\hline $\bolds{m}$ & \multicolumn{1}{c}{$\bolds{\kappa=1.0}$} &
\multicolumn{1}{c}{$\bolds{\kappa=1.5}$}
 & \multicolumn{1}{c@{}}{$\bolds{\kappa=2.0}$} \\
\hline
0 & 0.1\% & 0.0\% & 0.0\% \\
1 & 12.2\% & 2.6\% & 0.0\% \\
2 & 30.1\% & 10.9\% & 3.7\% \\
3 & 34.5\% & 17.5\% & 5.4\% \\
4 & 18.3\% & 23.6\% & 11.8\% \\
5 & 3.9\% & 12.9\% & 12.5\% \\
6 & 0.9\% & 31.9\% & 58.9\% \\
7 & 0.0\% & 0.6\% & 7.1\% \\
$>\!7$ & 0.0\% & 0.0\% & 0.6\% \\[3pt]
$p=1$ & 99.9\% & \multicolumn{1}{c}{100\%} & \multicolumn{1}{c}{100\%} \\
$p=0$ & 0.1\% & 0.0\% & 0.0\% \\
\hline
\end{tabular*}
\end{table}

\begin{figure}[t]

\includegraphics{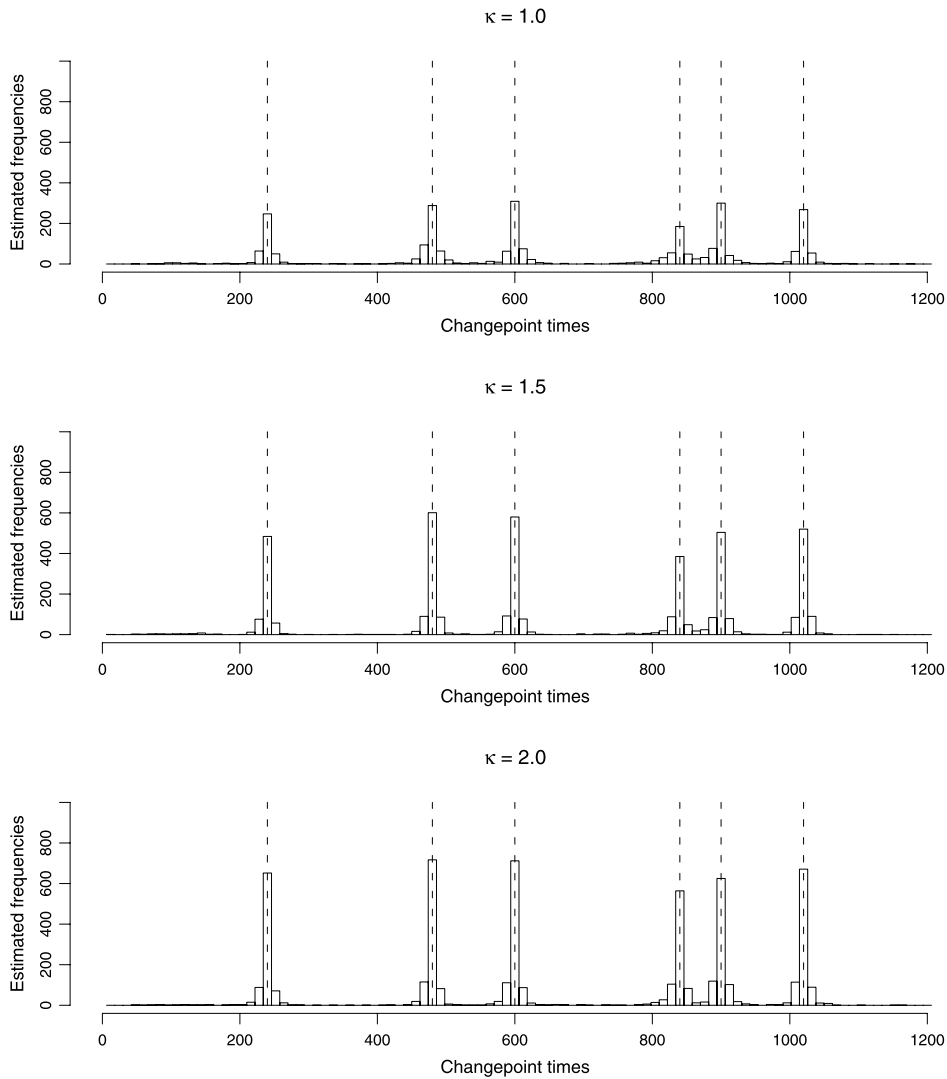}

\caption{Histograms of estimated changepoint
times.}\label{fig2}\vspace*{-6pt}
\end{figure}

Table \ref{tab2} and Figure \ref{fig2} summarize the results of the
simulations. Table \ref{tab2} reports empirical frequency distributions
of the number of estimated changepoints. Observe that the true value of
six changepoints is obtained more frequently as $\kappa$ increases.
When $\kappa=2.0$, the percentage of simulations where the correct
number of changepoints is estimated is 58.9\%, which is better than the
corresponding 43.4\% reported in Caussinus and Mestre
(\citeyear{Caussinus04}) that applies to uncorrelated and
time-homogeneous settings (i.e., 100 years of annual data). In
fairness, we note that the equivalent sample size of our simulated
series (the number of independent data points with the same periodic
variances) translates to more than the 100 independent data points of
Caussinus and Mestre (\citeyear{Caussinus04}) (we will not quantify
equivalent sample sizes further here). The correct number of
changepoints is identified only~0.9\% when $\kappa=1.0$ [this, however,
is also slightly better than the corresponding result in Caussinus and
Mestre (\citeyear{Caussinus04})]. It is clear that changepoint numbers
are underestimated in settings with relatively small $\kappa$. In fact,
the empirical mean (standard deviation in parentheses) of the
distributions in Table \ref{tab2} are 2.74 (1.07) for $\kappa=1.0$,
4.34 (1.48) for $\kappa=1.5$, and 5.413 (1.20) for $\kappa=2.0$.
Overall, one sees that changepoint shift sizes are critical in
changepoint detection, that the detection situation is difficult when
$\kappa$ is small, but that methods work reasonably well when~$\kappa$
is relatively large. Using monthly data (as opposed to annual averages)
also seems to improve changepoint detection power.

As for where the changepoints are estimated to occur, Figure \ref{fig2}
shows histograms of the estimated changepoint locations, reporting the
total number of times a changepoint is signaled at time $t$ for $1 \leq
t \leq N$ in the 1000 simulations. Observe that the histograms have
modes around the actual changepoint times. It is also evident that the
changepoints at times 840 and 900 were the most difficult to detect, a
feature attributed to the close proximity of the times of these two
changepoints (with the fifty-fifty up/down mean shift randomization
employed, the sign of these two mean shifts differ with probability
$1/2$, in which case their detection is relatively more difficult).

%
%t3 ###
\begin{table}[b]
\caption{Estimated number of changepoints for $n=100$}\label{tab3}
\begin{tabular*}{173pt}{@{\extracolsep{\fill}}ld{2.3}d{2.3}d{2.3}@{}}
\hline
%& \multicolumn{3}{|c|}{$a$ }\\ \cline{2-4}
$\bolds{m}$ & \multicolumn{1}{c}{$\bolds{a=1.0}$} & \multicolumn{1}{c}{$\bolds{a=2.0}$} & \multicolumn{1}{c@{}}{$\bolds{a=3.0}$} \\
\hline
0 & 6.2\% & 0.1\% & 0.0\% \\
1 & 29.5\% & 0.3\% & 0.0\% \\
2 & 32.5\% & 4.2\% & 0.0\% \\
3 & 22.8\% & 5.4\% & 0.0\% \\
4 & 7.4\% & 35.3\% & 7.7\% \\
5 & 1.5\% & 33.0\% & 4.4\% \\
6 & 0.1\% & 20.9\% & 81.7\% \\
7 & 0.0\% & 0.8\% & 6.1\% \\
$>\!7$ & 0.0\% & 0.0\% & 0.1\% \\
\hline
\end{tabular*}
\end{table}

Note that the correct autoregressive order $p=1$ was obtained virtually
all of the time. Hence, the time series model selection component seems
to be working well. As changing the trend parameter did not appreciably
affect results, we will not report separate tables with nonzero trends.

We now compare the MDL penalty more closely with the Caussinus--Lyazrhi
penalty used in Caussinus and Mestre (\citeyear{Caussinus04}). The
Caussinus--Lyazrhi penalty is larger than AIC or BIC penalties, but
does not penalize parameters in the mean function or consider
autocorrelation aspects. To make this comparison, 1000 series of length
100 were simulated with six changepoints always occurring at the times
20, 40, 50, 70, 75, and 85. The mean shift size parameter $\kappa$ was
changed to the parameter $a$ in Caussinus and Mestre
(\citeyear{Caussinus04}) to mimic their simulations. The errors in the
model were assumed to be Gaussian and independent. Note that the level
of changepoint activity relative to the sample size has increased
12-fold from the previous simulations. Table \ref{tab3} below lists
estimates of the relative frequencies of changepoints found by the
genetic algorithm with an MDL penalty when $\mu_\nu$ is held constant
with $\nu$. No trends were considered in this setup nor was tuning of
the genetic algorithm (varying its mutation probabilities, etc.)
considered in detail. The frequency distributions in Table \ref{tab3}
are approximately the same as those in Caussinus and Mestre
(\citeyear{Caussinus04}), perhaps slightly worse, in all cases. For
this sample size and level of changepoint activity, an MDL penalty
seems to perform about the same as the Caussinus--Lyazrhi penalty. Of
course, we reiterate that some gains are made by considering monthly
data in lieu of annual averages.

\begin{figure}[b]

\includegraphics{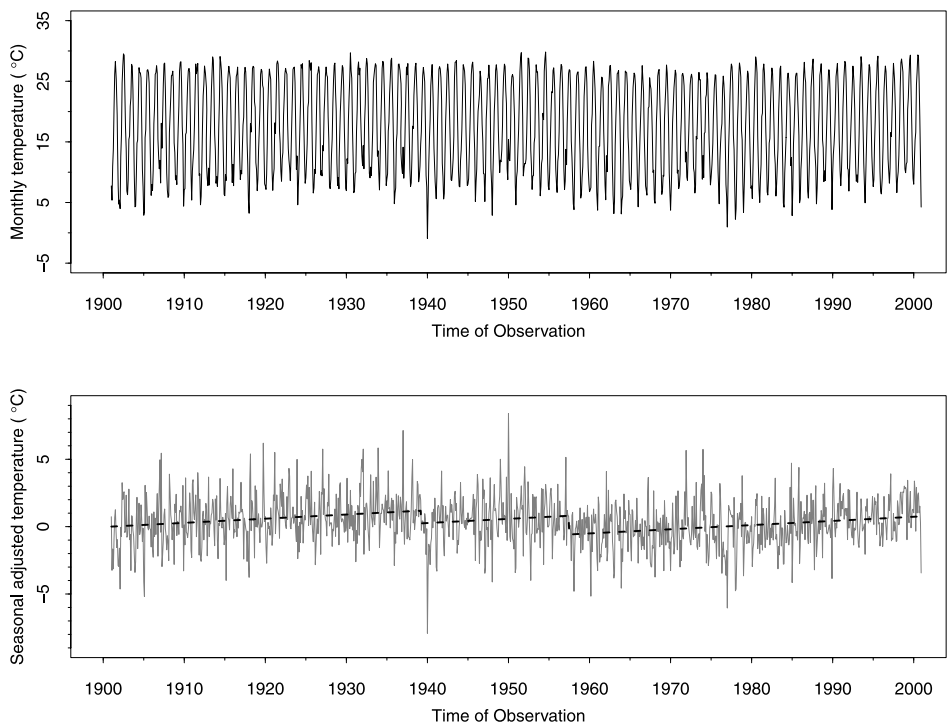}

\caption{The Tuscaloosa data with changepoint structure imposed.}
\label{fig3}
\end{figure}

%s6 ###
\section{The Tuscaloosa data}\label{sec6}

Figure \ref{fig3} plots a century of monthly data from Tuscaloosa,
Alabama recorded from January, 1901--December, 2000. A seasonal mean
cycle is visually evident in the data, but trends and mean shifts are
not readily apparent. Comparing the year-to-year jaggedness of the
seasonal throughs (the winter minimums) against the year-to-year
seasonal peaks (July maximums), it is discerned that this series has a
periodic variance with winter temperatures being much more variable
than summer temperatures. In fact, as we will see, the entire
autocorrelation structure of the series is periodic.

The Tuscaloosa series is one in which the station history is reasonably
documented. In particular, a catalog (called meta-data) exists that
notes the circumstances under which the data were recorded, including
the times of station relocations and instrumentation changes. This
said, meta-data files are notoriously incomplete [Menne and Williams
(\citeyear{Menne05})] and ``undocumented'' changepoints may lurk. The
Tuscaloosa series also has a moderately clean changepoint record with
only four major documented changes over a century of operation; as
noted before, the average United States temperature station experiences
about six changepoints per century [Mitchell (1953)]. In short, the
Tuscaloosa is a good ``proving ground'' series for changepoint methods.

Level shifts in temperature series are arguably the most important
factor in assessing temperature trends [see Lu and Lund
(\citeyear{Lu07})]. It has been argued in climate settings that the
manner in which changepoints are handled may be the most critical
factor in the global warming debate. Supporting this, temperature
insurance treaties on Wall Street are based solely on station location
and gauge properties, while ignoring long-term trends altogether.

Our methods were applied to the Tuscaloosa data. A reference series was
constructed by averaging three neighboring series located at Selma, AL,
Greensboro, AL, and Aberdeen, MS over the century of record. We work
with one reference series that averages three neighboring series to
expedite the discourse; methods that analyze all ${4 \choose2}$ pairs
of stations are discussed in Menne and Williams
(\citeyear{Menne05,Menne09}).

First, we examine the Tuscaloosa series without a reference. The fitted
MDL model has two changepoints at times 460 (April, 1939) and 679
(July, 1957). The mean function induced by these two changepoints, less
the seasonal cycle but including the trend, is plotted against the data
in Figure \ref{fig3}. The mean shift magnitudes of the 1939 and 1957
changepoints, in degrees Celsius, are both negative: $\hat{\Delta}_2=
-0.94 \pm0.20$ and $\hat{\Delta}_3= -2.33 \pm0.30$. The estimated trend
parameter is $\hat{\alpha}=0.00258 \pm0.00039$. The standard errors
were estimated from the fitted time series regression model with
generalized weighted least squares techniques. The estimated order of
the PAR model is $p=1$; a consequence of this is that the
autocovariance structure in the errors of the fitted models is indeed
periodic.

\begin{figure}[b]

\includegraphics{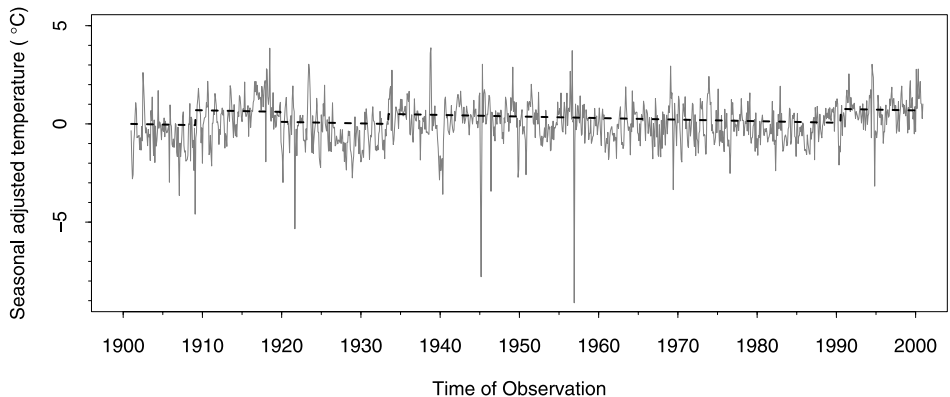}

\caption{The Tuscaloosa minus the reference data with changepoint
structure imposed.}\label{fig4}
\end{figure}

Second, we examine the Tuscaloosa minus the reference series. This
seasonally adjusted difference series is plotted in Figure \ref{fig4}.
In this target minus reference, four changepoints are flagged: March
1909, December 1919, July 1933, and August~1990. The estimates of the
$\Delta_i$'s are $\hat{\Delta}_2= 0.76 \pm0.11$, $\hat{\Delta}_3= 0.26
\pm0.11$, $\hat{\Delta}_4= 0.77 \pm0.13$, and $\hat{\Delta}_5= 1.47 \pm
0.19$. Note that a mean shift with the small magnitude of 0.26 has been
flagged. The trend estimate is $\hat{\alpha}= -0.00066 \pm0.00015$ and
the selected order of the autoregression is $p=1$. Observe that the
fitted order of the autoregression did not reduce from that for the raw
series; that is, periodic autocorrelation still exists in the target
minus reference series. Also, the trend for the target minus reference
series appears to be significantly negative. We comment that the two
large negative values occurring in the 1940s and the 1950s appear to be
decimal typos in the raw data; that is, the monthly average temperature
for Tuscaloosa was entered as ten degrees too small. We make this claim
after examining additional reference series from various cities close
to Tuscaloosa. Whereas the series in this database have been quality
checked to some degree, errors like this may still exist. We reran the
analysis above after replacing these two values by (1) their estimated
seasonal means $\hat{\mu}_\nu$ and (2) what we believe are the correct
values, that is, adding 10 degrees to both outliers. In both cases,
four changepoints with similar magnitudes and times to the ones above
are found. The 1957 changepoint flagged in the target series has not
been flagged in any version of the target minus difference series. The
1909 changepoint is possibly attributed to a changepoint in the
reference series: Greensboro reports a time of observation change in
1906 and Aberdeen reports a station relocation in 1915.

The meta-data show four changepoints in this series: the first was a
station relocation in November of 1921, the second was a station
relocation in March of 1939, the third was a station relocation during
June of 1956 and an accompanying instrumentation change in November of
1956 (we regard this as one changepoint), and the fourth is a station
relocation and instrumentation change in May of 1987. The reference
series analysis seems to have correctly identified three of these four
changepoints (we are liberally including the 1933 flagged changepoint
time as correctly identifying the 1939 changepoint), missing the 1956
changepoint and adding a 1909 changepoint. The raw target series
analysis misses the 1921 and 1987 changepoints, but finds the 1956
changepoint; also, the estimated time of the 1939 changepoint is much
closer to its true value than that for the reference analysis. Overall,
it seems that the reference analysis is superior to simple target
series analysis, but that one can learn something with both analyses.

\begin{figure}

\includegraphics{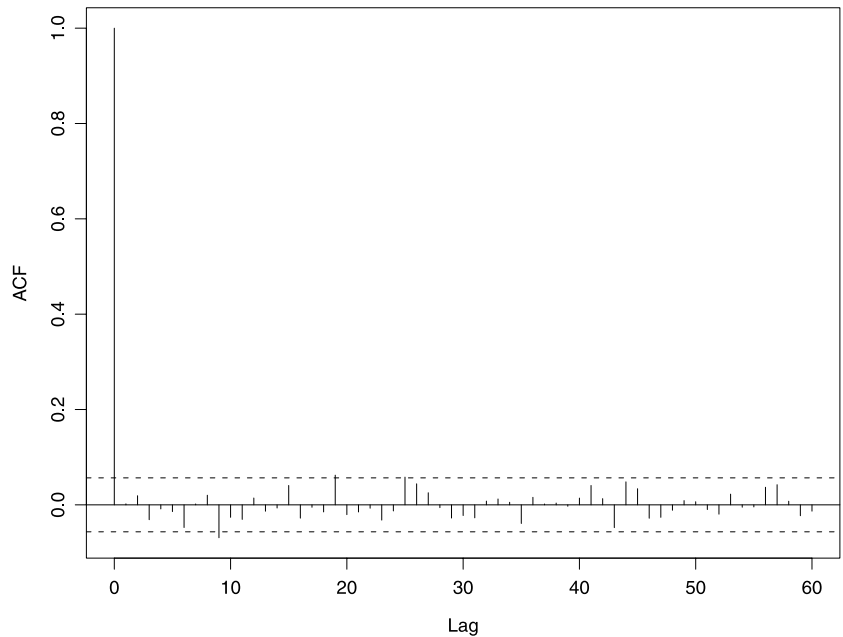}

\caption{Sample autocorrelations of target series residuals.}
\label{fig5}
\end{figure}

\begin{figure}

\includegraphics{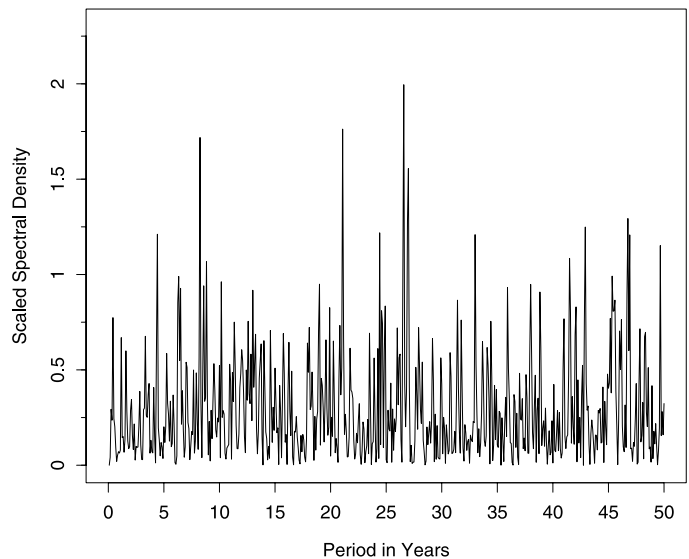}

\caption{Periodogram of target minus reference series residuals.}
\label{fig6}
\end{figure}

We caution the reader that trends in some monthly temperature series,
especially when the series is aggregated over a large geographic
region, may not be well described by a linear regression component. As
noted by Handcock and Wallis~(\citeyear{Handcock94}), trends at
localized series are more likely to be adequately described with a
simple linear structure. As a final diagnostic check, residuals from
the model fits were computed. Figure \ref{fig5} shows the sample
autocorrelation of the residuals for the target series over the first
60 lags. The dashed lines are 95\% pointwise confidence bounds for
white noise. As only three of the sample autocorrelations lie outside
the bounds (and then only slightly so), the model appears to have
fitted the data well. Figure \ref{fig6} shows the periodogram of the
residuals from the target minus reference series. A long memory
structure is not readily evident in this plot.

%s7 ###
\section{Comments}\label{sec7}

Time series parsimony may be an issue with periodic data. Specifically,
the penalty in (\ref{piece2}) essentially assumes that the PAR(1) model
requires $(p+1)T$ distinct parameters. In practice, changes in climate
processes from season to season are slow/smooth. Low order Fourier
series expansions, such as those in Lund, Shao, and Basawa
(\citeyear{Lund05}), can statistically simplify the model and serve to
lessen the penalty for time series components. This issue is likely to
be paramount should daily data be considered.

\section*{Acknowledgments} Comments from two referees
and the editor substantially improved this paper.
%--------------------------------------------------------------

%

%

\printaddresses

\end{document}